\documentclass[pre,twocolumn,superscriptaddress,showpacs]{revtex4}
\usepackage{epsfig,amsmath,amssymb,graphicx,color,calc}
\newcommand{\be}{\begin{equation}}
\newcommand{\ee}{\end{equation}}
\renewcommand{\phi}{\varphi}
\newcommand{\ba}{\begin{eqnarray}}
\newcommand{\ea}{\end{eqnarray}}

\begin{document}

\title{Scaling of the glassy dynamics of
soft repulsive particles: a mode-coupling approach}

\author{Ludovic Berthier}
\affiliation{Laboratoire des Collo{\"\i}des, Verres
et Nanomat{\'e}riaux, UMR CNRS 5587, Universit{\'e} Montpellier 2,
34095 Montpellier, France}

\author{Elijah Flenner}
\affiliation{Department of Chemistry, Colorado 
State University, Fort Collins, CO 80523}

\author{Hugo Jacquin}
\affiliation{Laboratoire Mati\`ere et Syst\`emes Complexes, UMR CNRS 7057,
Universit\'e Paris Diderot -- Paris 7, 10 rue Alice Domon et L\'eonie 
Duquet, 75205 Paris cedex 13, France}

\author{Grzegorz Szamel}
\affiliation{Laboratoire des Collo{\"\i}des, Verres
et Nanomat{\'e}riaux, UMR CNRS 5587, Universit{\'e} Montpellier 2,
34095 Montpellier, France}
\affiliation{Department of Chemistry, Colorado State University, 
Fort Collins, CO 80523}

\date{\today}
 
\begin{abstract}
We combine the hyper-netted chain approximation of liquid state theory with 
the mode-coupling theory of the glass transition to 
analyze the structure and dynamics of
soft spheres interacting via harmonic repulsion. We determine the locus 
of the fluid-glass dynamic transition in a 
temperature -- volume fraction 
phase diagram. The zero-temperature (hard sphere) glass transition
influences the dynamics at finite temperatures in its vicinity. This directly 
implies a form of dynamic scaling 
for both the average relaxation time and 
dynamic susceptibilities quantifying dynamic heterogeneity.
We discuss several qualitative disagreements between theory and 
existing simulations at equilibrium. Our theoretical results are, however,  
very similar to numerical results for the driven athermal dynamics 
of repulsive spheres, suggesting that `mean-field' mode-coupling 
approaches might be good starting points 
to describe these nonequilibrium dynamics.
\end{abstract}

\pacs{05.20.Jj, 64.70.qd}


\maketitle

\section{Introduction}

An assembly of hard spherical particles undergoing Brownian motion,
if it can avoid crystallization (\textit{e.g.} due to polydispersity),
at a sufficiently high volume fraction
undergoes a glass transition to an amorphous solid state \cite{pusey}. 
In experiments, structural relaxation of
colloidal hard sphere systems stops  near volume fraction 
$\phi \approx 0.60$~\cite{luca}.  
The phenomenology of this so-called colloidal glass transition is strikingly 
reminiscent of the molecular glass transition observed 
upon decreasing the temperature in glass-forming liquids.
When Brownian motion is negligible, a hard sphere system 
undergoes instead a jamming transition near $\phi \approx 0.64$; 
it acquires rigidity 
by building a mechanically stable 
network of contacts between particles~\cite{bernal}. 
This is most readily observed in granular materials. 

While the glass and jamming transitions of hard sphere systems 
have been widely studied for a long time,  the study
of dense systems composed of soft 
repulsive particles is, by comparison, in its infancy.
Colloidal particles with tunable softness are now 
routinely prepared in the laboratory~\cite{yodh,trappe,maret,weitz} and 
examples of compressible grains abound~\cite{peas,foams,tapioca}.
The rheological, structural, or dynamical properties of both types of systems
are currently actively studied experimentally
by several groups~\cite{yodh,trappe,maret,weitz,peas,foams,tapioca}. This 
justifies current theoretical efforts to understand the behaviour of 
dense assemblies of soft repulsive particles 
both at finite temperatures~\cite{tom2,tom,liu,xu}, 
relevant for colloidal systems,
and in the zero-temperature limit relevant for granular 
materials~\cite{durian,ohern,teitel,claus,hatano}.

In this paper, we combine the hyper-netted chain approximation of 
liquid state theory~\cite{barrat} with 
mode-coupling theory~\cite{gotze} to analyze the equilibrium 
structure and dynamics of dense systems of soft particles
with finite range, harmonic repulsion~\cite{durian}:
\be
V(r < \sigma) = \epsilon (1-r/\sigma)^2,
\label{pair}
\ee
where $\epsilon$ determines the strength of the repulsion, 
$\sigma$ is the particle diameter, and 
$r$ the distance between two particles. 
Particles separated by $r>\sigma$ do not interact, 
$V(r>\sigma)=0$.
The control parameters for this system are therefore the 
volume fraction $\phi = \pi \sigma^3 \rho / 6$, with $\rho$ the number density, 
and the ratio of the temperature and $\epsilon$. In the following we use reduced units:
we give lengths in units of $\sigma$ and temperature in units of $\epsilon$.

The system of harmonic spheres was originally introduced in the context of 
the zero-temperature jamming transition~\cite{durian}. More recently, its behaviour 
was investigated at finite temperature 
using molecular dynamics computer simulations~\cite{tom2,tom,liu}. 
To our knowledge, 
this model system was not studied theoretically in the context of liquid 
state and mode-coupling theories. However, its physics should be similar to a number of 
similar models such as Hertzian spheres~\cite{frenkel} 
or the Gaussian core model~\cite{stillinger,louis}. 

In both thermal and athermal contexts 
various scaling relations were reported for the dynamics of harmonic spheres 
in the vicinity of the glass or jamming transitions 
occurring in the hard sphere 
limit~\cite{teitel,hatano,tom2,tom}.
This type of scaling is usually rationalized by postulating that
harmonic spheres in fact behave as an `effective' 
hard sphere system with a `renormalized' volume fraction~\cite{tom2}. 
However, the scaling formulae are typically established using largely empirical 
procedures. Our primary goal
in this work is to derive this scaling behaviour using a liquid 
state theoretical approach in order to put it on a firmer basis, 
at least for systems in thermal equilibrium.

The paper is organized as follows. We present 
theoretical methods which we use to obtain the structure and 
dynamics of harmonic spheres in Sec.~\ref{theory}. 
We then present results for the equilibrium dynamics in 
Sec.~\ref{eqscaling}. We describe the phase diagram in Sec.~\ref{phasediag}.
In Sec.~\ref{chi4} we analyze dynamic heterogeneity using 
three-point dynamic susceptibilities.
We conclude the paper in Sec.~\ref{conclusion}. 

\section{Theoretical approach}
\label{theory}

Our theoretical approach consists of two steps. 
First, we use the hyper-netted chain (HNC) approximation
of liquid state theory to obtain the static structure of the 
harmonic sphere system. Next, we use the predicted static structure factor 
as input of the mode-coupling equations to obtain time correlation functions
of harmonic spheres and we use these functions to determine the 
dynamical behaviour. 
In the next two subsections we briefly describe the two elements 
of our approach. 

\subsection{HNC equations and static behaviour}
\label{sectionhnc}

The HNC approximation \cite{barrat} results in a closed equation 
for the pair correlation 
$g(r)$ of a fluid. It reads: 
\be
g(r) = \exp[ -\beta V(r) + g(r) - 1 -c(r) ],
\label{hnc}
\ee
where $\beta = 1/T$ and $c(r)$ is the direct correlation function defined 
through the Ornstein-Zernike equation:
\be
g(r)-1 = c(r) + \rho \int dr' c(|r-r'|) [g(r')-1].
\label{OZ}
\ee
We solve Eq.~(\ref{hnc}) numerically using an iterative
procedure. Since the direct correlation function is smoother 
than the pair correlation function, the HNC equation 
is more easily solved in terms of $c(r)$.
Given the solution for $c(r)$ after $i-1$ iterations, $c_{i-1}(r)$,
we obtain 
the solution at step $i$ as follows: 
\begin{align} 
& c_{i-1}(r) \xrightarrow{\rm FT} \hat{c}_{i-1}(q) \xrightarrow{\rm OZ} 
\hat{g}_{i-1}(q) \xrightarrow{{\rm FT}^{-1}} g_{i-1}(r) \nonumber \\
& \xrightarrow{\rm HNC} {c}(r) 
\rightarrow c_i(r) = \alpha c(r)+ (1-\alpha) 
c_{i-1}(r),
\label{iteration}
\end{align}
where the first and third steps are Fourier transforms
performed using fast Fourier transforms,
the second one uses the Ornstein-Zernike relation (\ref{OZ}), 
the fourth one uses the HNC closure relation (\ref{hnc}), and the 
last step involves combining the new direct correlation function $c(r)$ 
with $c_{i-1}(r)$ using a mixing parameter $\alpha$. 
Convergence is achieved when the difference between 
$c_i(r)$ and $c_{i-1}(r)$ in Eq.~(\ref{iteration}) 
becomes smaller than some 
prescribed precision. 

Since we want to access very low temperatures where the pair potential 
in Eq.~(\ref{pair}) becomes equivalent to a hard sphere potential, 
some attention must be paid to the 
discretization scheme we use. 
Additionally, the fluid develops medium-range 
structure when density increases and/or temperature decreases, so that 
the cut-off in real space must be large enough.
We used a real space cutoff $L=32$, and 
discretized the interval $[0,L]$  using $2^n$ points, which is  
convenient for the fast Fourier transform algorithm. 
To properly represent the low temperature 
behaviour, discretization must be accurate enough that the 
factor 
\be
Y(r,T) \equiv \exp[-\beta (1-r)^2]
\ee 
is correctly described even at very low $T$. We note that, for a 
given discretization, there necessarily exists 
a temperature, $T_n$, below which $Y(r,T)$ is not well-described.
It is easy to see that this temperature scales as $T_n \sim 2^{-2n}$.

\begin{figure}
\includegraphics[width=8.5cm]{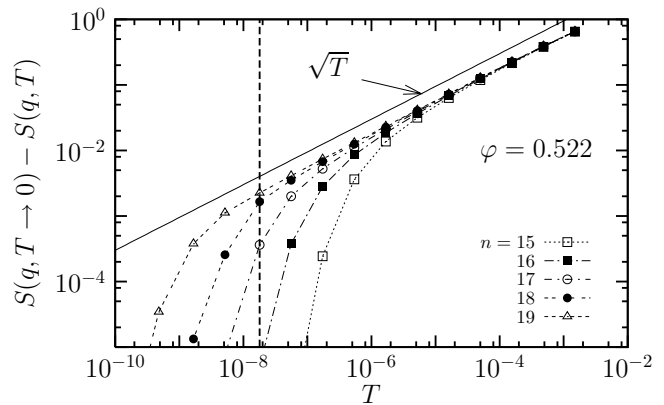}
\caption{\label{figconvergence} Convergence of the numerical 
solution of the HNC structure factor to its $T=0$ hard sphere limit
for $q$ near the first diffraction peak at $\phi=0.522$.
For a given number $2^n$ of discretized values of $r$ used to solve 
Eq.~(\ref{hnc}), there exists a temperature above which 
the $\sqrt{T}$ convergence in Eq.~(\ref{scalingT}) is obeyed.
The vertical line shows the lowest value of $T$ used in this work.}
\end{figure}

This behaviour is ilustrated in Fig.~\ref{figconvergence} which 
shows how the structure factor $S(q,T)$
of harmonic spheres at $\phi=0.522$ reaches its $T \to 0$ value
for increasingly finer discretizations (data are shown for $q$
near the first diffraction peak).
To get accurate results down to $T=10^{-8}$, as shown below, we 
need to use $n > 18$. The value $n=19$ is used throughout this 
paper. Additionally, since we need structure factors 
for temperatures and densities in very narrow ranges near 
the dynamic singularities, we had to pay close attention to the accuracy
of the numerical solution of the HNC equation in order to resolve very close 
state points.

In Fig.~\ref{figconvergence} we also show that difference between the 
finite temperature $S(q,T)$ and its hard sphere ($T=0$) limit  
scales as $\sqrt{T}$. This behaviour can be derived
as follows. Since $g(r<1,T=0)=0$ for hard spheres we can decompose
the difference $S(q,T) - S(q,0)$ as follows
\begin{align}
S(q,T) - S(q,0) & =  4 \pi \rho \int_0^1 dr r^2 \frac{\sin qr}{qr}
g(r,T) \nonumber \\
& +  4 \pi \rho 
\int_1^\infty dr r^2 \frac{\sin qr}{qr} [g(r,T)-g(r,0)]. \nonumber
\end{align}
Using the fact that in the low temperature $T \to 0$ limit 
$Y(r,T)$ becomes a narrow peak located at $r=1$, it
is possible to approximate the  
first term in this expression by
\be
2 \pi^{3/2} \rho y(r=1,T=0) \frac{\sin q}{q} \sqrt{T}. 
\label{sqrtT}
\ee
where $y(r,T) = g(r,T)/Y(r,T)$ is the cavity function. 
Since both $g(r,T)-g(r,0)$ and $S(q,T)-S(q,0)$ are of the same order,  
we conclude that 
\be
S(q,T) - S(q,0) \sim \sqrt{T},
\label{scalingT}
\ee 
and this scaling is satisfied, see Fig.~\ref{figconvergence}.
The $\sqrt{T}$ temperature dependence of the difference 
$S(q,T) - S(q,0)$ will play a crucial role 
in the next section, Sec. \ref{eqscaling}. We will use it to motivate 
the dynamic scaling in the vicinity of the hard-sphere transition, 
within the mode-coupling approximation.

In a recent molecular dynamics investigation \cite{tom2}, 
the temperature dependence of the 
energy density, $e(T)$, was used to relate soft spheres to hard particles. We 
can use a similar reasoning to predict the low temperature behaviour of $e(T)$.
The energy density can be expressed in terms of the pair correlation function  
\be
e(T) = 2\pi \rho \int_0^1 dr r^2 V(r) g(r,T).
\ee
The convergence of $e(T)$ to its $T=0$ value 
$e(T=0)=0$ can be estimated by 
making explicit the $Y(r,T)$ 
factor in $g(r,T)$ and using the cavity function.
In this way we obtain 
\be 
e(T) \approx  \frac{ \pi^{3/2}}{2} \rho y(r=1,T=0) T^{3/2}.
\label{T32}
\ee
Molecular dynamics investigation~\cite{tom2} reported a 
power law behaviour of the energy density, $e(T) \sim T^\mu$, with exponent 
$\mu$ crossing over from  $\mu =3/2$ at low volume fraction
to $\mu \approx 1.3$ at larger volume fraction. This indicates that the low 
temperature scaling regime (\ref{T32}) 
is not accessible at large densities in molecular dynamics simulations.

\subsection{MCT analysis and dynamic behaviour}
\label{grzegorz}

The mode-coupling theory (MCT) \cite{gotze} was originally derived to describe
the dynamics of Newtonian systems \cite{bengtzelius}. An analogous 
theory was later derived for Brownian systems \cite{sl}. Here we briefly
present the latter version of MCT.  

The starting point of the theory is an exact equation for the time 
derivative of the intermediate scattering function $F(q;t)$ in terms
of the so-called irreducible memory function,
\be\label{FMirr}
\partial_t F(q;t) = -\frac{D_0 q^2}{S(q)} F(q;t) 
- \int_0^t dt' M^{\mathrm{irr}}(q;t-t') \partial_{t'} F(q;t').
\ee
Here $D_0$ is the diffusion coefficient of an isolated Brownian particle.
Irreducible memory function $M^{\mathrm{irr}}(q;t)$ can be expressed in
terms of a time-dependent four-point density correlation function evolving
with the so-called irreducible dynamics. The main approximation of MCT
consists in factorizing this four-point function. In this way this somewhat
mysterious quantity is reduced to a product of two intermediate scattering functions.
After an additional technical approximation (which was independently shown to
be quite innocuous) one arrives with the following expression for the 
irreducible memory function
\begin{eqnarray}\label{Mirr}
&& M^{\mathrm{irr}}(q;t) =
\frac{\rho D_0}{2q^2} \int \frac{d\mathbf{q}_1}{(2\pi)^3}
\left[\mathbf{q}\cdot\mathbf{q}_1 \hat{c}(q_1) \right.
\\ \nonumber && + \left.
\mathbf{q}\cdot(\mathbf{q}-\mathbf{q}_1) \hat{c}(|\mathbf{q}
-\mathbf{q}_1|)\right]^2
F(q_1;t)F(|\mathbf{q}-\mathbf{q}_1|;t).
\end{eqnarray}

Equations~(\ref{FMirr}-\ref{Mirr}) allow us to evaluate the time dependence of the
intermediate scattering function. The only input required is the static structure 
factor $S(q)$. It enters into Eqs. (\ref{FMirr}-\ref{Mirr}) (note that
$\hat{c}(q) = (1-1/S(q))/\rho$, from Eq.~(\ref{OZ})) 
and it also provides the initial condition
for the intermediate scattering function, $F(q;t=0)=S(q)$. It is easy to see that 
the natural time unit for our system of harmonic spheres is $\sigma^2/D_0$.
In the following all times are given in terms of this unit. 

\begin{figure}
\includegraphics[width=8.5cm]{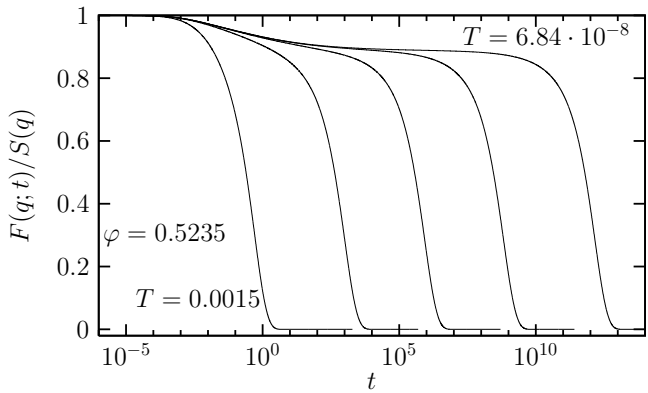}
\caption{\label{fqt} Time dependence of the normalized 
intermediate scattering function $F(q;t)/S(q)$ as predicted by 
the mode-coupling theory for harmonic spheres at $\phi=0.5235$,
for wavevector corresponding the  first peak in the static structure 
factor, $q=q_{\mathrm{max}}\approx 7.17$. 
The static structure factor used as input in mode-coupling equations was
obtained from the hyper-netted chain approximation. The lines correspond to
the following temperatures (from left to right): 0.0015, $1.61\times 10^{-5}$,
$5.39\times 10^{-7}$, $1.00\times 10^{-7}$, and $6.84\times 10^{-8}$.}
\end{figure}

Numerical solution of Eqs.~(\ref{FMirr}-\ref{Mirr}) is somewhat 
complicated because
one needs to describe evolution of the intermediate scattering function on
very widely separated time scales (see Fig.~\ref{fqt}). The commonly used algorithm
was first described in Ref.~\cite{fuchs}; here we use the implementation described 
in considerable detail in Ref. \cite{flennerszamel}. Briefly, 
the basic steps to the algorithm are as follows. The integro-differential 
equation is discretized and solved for $2 N_s$ steps with
a finite time step of $\delta t$ using any suitable numerical algorithm.
After $2 N_s$ steps are complete, the time step is doubled and the
results from the initial $2 N_s$ steps are mapped into a new equally
spaced set of $N_s$ values for the quantities needed to continue the
numerical algorithm. This mapping includes the integrals as well as 
the intermediate scattering functions. Then
the numerical algorithm is restarted with the new time step and
continued for another $N_s$ time steps and the mapping is performed 
again. This procedure is continued until a convergence condition is satisfied.
In the present work we used 300 equally spaced wave-vectors with spacing 
$\delta\approx 1.96$, the first wave-vector at $k_0=\delta/2$ and the
largest wavevector at $k_{\mathrm{max}}\approx 58.81$. 

In Fig. \ref{fqt} we show the prediction of the mode-coupling theory for the 
time dependence of the normalized  
intermediate scattering function $F(q;t)/S(q)$ for harmonic spheres at $\phi=0.5235$. 
With decreasing temperature the relaxation becomes progressively slower.
One should notice that an intermediate time plateau is developing which is
the manifestation of the cage effect: with decreasing temperature particles are 
trapped longer and longer within their first solvation shells and the final
($\alpha$) relaxation shifts to longer and longer times. MCT
predicts that at this volume fraction at temperature 
$T_c(\phi=0.5235)\approx 6.594\times 10^{-8}$ the plateau extends to infinite times
and the system undergoes an ergodicity breaking transition which is commonly
referred to as the glass transition.

In this work we are primarily focused on the temperature and volume fraction
dependence of the $\alpha$ relaxation time which we define in the standard 
way, $F(q_{\mathrm{max}};\tau_{\alpha}) = e^{-1}$, where $q_{\mathrm{max}}$ is the 
position of the first peak in the static structure factor. 
MCT makes a number of detailed predictions for various aspects of the time
dependence of intermediate scattering functions (including power law approach to 
and departure from the intermediate time plateau, and the wavevector
dependence of the relaxation time). The investigation of the temperature and volume 
fraction dependence of these properties are left for future work.

The hard sphere glass transition within MCT is usually investigated using
as the input the static structure factor obtained from the Percus-Yevick (PY)  
approximation. For the discretization used in this work MCT combined with the
PY structure factor predicts the glass transition at 
$\phi^{\mathrm{PY}}_c\approx 0.5159$. Moreover, MCT predicts that upon 
approaching $\phi^{\mathrm{PY}}_c$ the $\alpha$ relaxation time diverges
algebraically, 
$\tau_{\alpha} \sim (\phi^{\mathrm{PY}}_c-\phi)^{-\gamma^{\mathrm{HS,PY}}}$,
with the exponent $\gamma^{\mathrm{HS,PY}}\approx 2.59$. 

In this work we use the hyper-netted chain approximation for the static structure
factor because we are mostly interested in effects of finite temperature. 
However, we anticipate that the low temperature results will be influenced by
the behavior of the hard sphere system. Therefore, we 
also solved MCT 
equations in the $T \to 0$ limit using as input the
structure factor predicted by the 
hyper-netted chain approximation in this limit. 
For the discretization used in this work MCT 
combined with the HNC structure factor predicts the glass transition at 
$\phi_c\approx 0.52315$. Moreover, according to MCT upon approaching this
critical volume fraction the $\alpha$ relaxation time diverges
algebraically, 
\be
\tau_\alpha^{\rm HS}(\phi) \approx \frac{\tau_0}{(\phi_c - 
\phi)^{\gamma^{\rm HS}}}
\label{T=0}
\ee 
with the exponent $\gamma^{\rm HS} \approx 3.26$. 

It is well known that in real colloidal systems the glass transition
predicted by mode-coupling theory is avoided.
Typically, as shown in recent contributions \cite{luca,luca2,tom},
one can find an intermediate range of volume fractions in which the 
volume fraction dependence of the $\alpha$ relaxation time can be fitted to a power
law with the exponent close to that predicted by MCT (one should note that when 
this procedure is used the critical volume fraction is one of the fitting parameters; 
typical values obtained from fits are about 10\% different from MCT predictions). 
The result is that MCT-predicted power law divergence of the $\alpha$ relaxation time 
describes well 
the experimental and simulational data over approximately 3 decades
of $\tau_{\alpha}$. Recent experimental and simulational results reported in 
Refs.~\cite{luca,luca2,tom} suggest that upon increasing volume fraction further
this approximate power law is
followed by an `activated' regime according to which the $\alpha$
relaxation time has an essential 
singularity divergence at a higher volume fraction.
 
\section{Dynamic scaling at equilibrium}
\label{eqscaling}

The tools described in the previous section 
allow us to obtain, for any given state point 
$(\phi,T)$, the relaxation time $\tau_\alpha(\phi,T)$ of the 
harmonic sphere system within the MCT approximation.
We will report results for a broad range 
of volume fractions, $\phi \in [0.51, 0.90]$, 
and temperatures, $T \in [10^{-8}, 10^{-2}]$. 

\begin{figure}
\includegraphics[width=8.5cm]{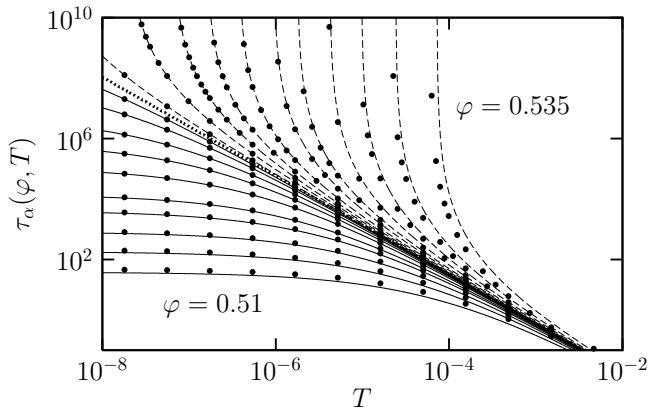}
\caption{\label{fig1} Relaxation time of harmonic spheres
as a function of temperature for various volume fractions 
in the vicinity of the hard sphere glass transition. 
The lines are the analytical formula (\ref{scaling}) 
using the scaling functions in Eq.~(\ref{fpm}); full lines correspond
to $\phi< \phi_c$, dashed lines to $\phi>\phi_c$. 
The power law in Eq.~(\ref{Tphic}) for $\phi = \phi_c \approx 0.52315$ 
is shown with a dotted line. 
Corrections to scaling are seen for $\phi \le 0.515$
and $\phi \ge 0.53$.
Volume fractions are: (i) 0.51, 0.515, 0.518, 0.52, 0.521, 0.522, 
0.5225, 0.5228, 0.523, 0.5231 (ii) 0.5232, 0.52334, 0.5235, 0.5237, 
0.524, 0.5245, 0.525, 0.526, 0.5275, 0.53, 0.535.
}
\end{figure}

In Fig.~\ref{fig1} we show the evolution of $\tau_\alpha(\phi,T)$ in 
the vicinity of the hard sphere glass transition 
occurring at $T=0$ and $\phi_c \approx 0.52315$. 
The behaviour observed at finite temperature 
is easily explained. When $\phi < \phi_c$, the relaxation time
increases when $T$ decreases, but it saturates at low 
temperature to its hard sphere value which is finite 
at these densities. When increasing the volume fraction 
closer to $\phi_c$, this hard sphere value becomes larger, and 
the low temperature limit is reached at a lower temperature. 
For $\phi > \phi_c$ the system is a hard sphere glass 
in the $T \to 0$ limit and so $\tau_\alpha(\phi,T \to 0) = \infty$.
It is clear, however, that the system hits a finite temperature 
singularity at a critical temperature, $T_c(\phi)$ which 
increases continuously from $T_c(\phi=\phi_c)=0$ when 
$\phi$ increases.

At this stage of the description, these data resemble 
the ones found in numerical simulations~\cite{tom2,tom}.
In Refs.~\cite{tom2,tom} a scaling analysis of the relaxation time 
was performed assuming that harmonic spheres at low temperature 
resemble an `effective' fluid of hard spheres. Physically, 
this means that the softness of the potential allows 
small overlaps between particles at low temperatures, so that
the effective radius of the particles is reduced by thermal
fluctuations. While the temperature dependence 
of the energy density was used to estimate the
average overlap, and in turn, the effective hard 
sphere diameter in Ref.~\cite{tom2}, 
in the context of the mode-coupling approach 
a different route should be used to map soft to hard particles.

Within MCT, at a fixed number density the dynamics of the system
is uniquely controlled by the evolution of the static
structure factor. This suggests that the most 
efficient way to map soft to hard particles is by matching 
the structure factor of harmonic spheres at $(\phi,T)$
to the one of a hard sphere system at an effective 
volume fraction $\phi_{\rm eff} = \phi_{\rm eff}(\phi,T)$. Combining 
the low temperature behaviour of $S(q,T)$ shown
in Eq.~(\ref{scalingT}) with the fact that $S(q)$ is 
smooth function of the volume fraction we easily find that 
\be
\phi_{\rm eff}(\phi,T) \approx \phi - c \sqrt{T},
\label{phieff}
\ee
where $c$ is a positive prefactor with subleading dependencies on 
temperature and volume fraction.

The discussion in the two previous paragraphs leads to the following 
form of dynamic scaling that should be obeyed by 
the relaxation time of the harmonic sphere system:
\be
\tau_\alpha(\phi,T) \approx \frac{\tau_0}{|\phi_c - 
\phi|^{\gamma^{\rm HS}}}
f_{\pm} \left( \frac{|\phi_c - \phi|}{\sqrt{T}} \right),
\label{scaling}
\ee
where the scaling functions $f_{\pm}(x)$ respectively 
refer to volume fractions above and below $\phi_c$. 
To be consistent with the qualitative behaviour described above, 
the scaling functions $f_\pm(x)$ must have 
the following limiting behaviors: to recover 
the hard sphere plateau at low $T$ below $\phi_c$ we need to have
\be
f_{-}(x \to \infty) \sim const ;
\label{C1}
\ee
whereas to have a well-behaved $\tau_\alpha(\phi,T>0)$ across $\phi_c$ we
have to require
\be
f_{-}(x \to 0) \sim f_{+}(x \to 0) \sim x^{\gamma^{\rm HS}}.
\label{C2}
\ee 
The latter limiting behavior implies that 
\be
\tau_\alpha(\phi=\phi_c,T) \sim \left( \frac{1}{T} 
\right)^{\gamma^{\rm HS}/2}.
\label{Tphic}
\ee
Finally, a finite temperature algebraic singularity 
is obtained for $\phi > \phi_c$ if  there exists some 
$x_\star$ such that
\be
f_{+}(x \to x_\star^-) \sim (x_\star -x)^{-\gamma^{\rm HS}}.
\label{C3}
\ee

\begin{figure}
\includegraphics[width=8.5cm,clip]{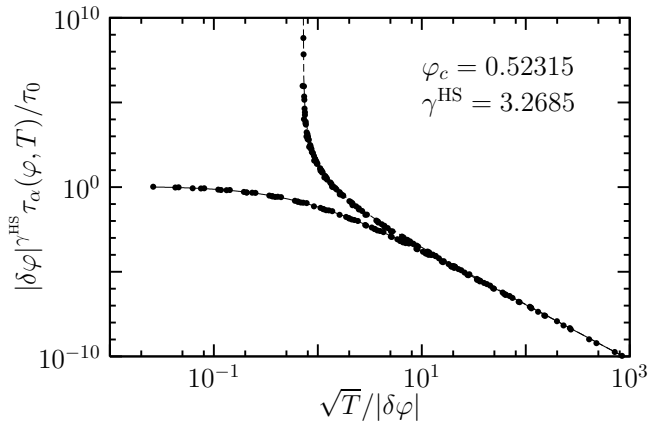}
\caption{\label{collapse} Collapse of the data shown 
in Fig.~\ref{fig1} using the scaled variables suggested 
by Eq.~(\ref{scaling}) for all data in the range 
$0.515 < \phi < 0.53$. This data collapse involves no free parameter.
The lines through the points represent the empirical 
scaling functions $f_{\pm}(x)$ in Eq.~(\ref{fpm}). 
}
\end{figure}

We have found excellent agreement of the MCT predictions with the 
scaling form in Eq.~(\ref{scaling}), which suggests 
that the relaxation times at various $\phi$ and $T$ 
can all be collapsed along two branches by plotting the rescaled time,
$|\delta \phi|^{\gamma^{\rm HS}} \tau_\alpha(\phi,T)$,
as a function of the rescaled distance to the critical point,
$|\delta \phi| / \sqrt{T}$,
where $\delta \phi \equiv \phi_c - \phi$.
The data collapse is presented in Fig.~\ref{collapse}.
It works remarkably well for the range
of volume fraction $0.515 < \phi < 0.53$. 
It should be noted that no free parameter is involved in 
this data collapse, which is uniquely controlled by the $T=0$ 
hard sphere results from Eq.~(\ref{T=0}), and by using the appropriate 
relation between temperature and density from 
Eq.~(\ref{phieff}). In Refs.~\cite{tom,tom2}, a similar data collapse
was used in the opposite direction  
to infer the hard sphere behaviour from the finite temperature 
dynamics of the harmonic sphere system, a philosophy which 
is clearly supported by the results presented in this section.
   
The next step is to take Eq.~(\ref{phieff}) more literally and to combine it
with the results for 
the dynamics of the hard sphere system discussed in Sec.~\ref{grzegorz}
to make the following ansatz
\be\label{ansatz}
\tau_\alpha(\phi,T) \approx 
\tau_\alpha^{\rm HS} [ \phi_{\rm eff}(\phi,T) ].
\ee
Eq.~(\ref{ansatz}) leads to the following scaling functions:
\be
f_{\pm} (x) = (\frac{1}{x} \mp a )^{-\gamma^{\rm HS}}, 
\label{fpm}
\ee
with $a \approx 0.72$ being the only adjustable numerical 
factor (note that $a$ is related to $c$ in Eq.~(\ref{phieff}))
since the values $\gamma^{\rm HS}=3.26$ and 
$\phi_c = 0.52315$ are directly taken 
from hard sphere results, while the scaling variable 
$x = |\delta \phi| / \sqrt{T}$ was derived in Sec.~\ref{sectionhnc}.
Clearly, these scaling functions are fully compatible 
with the constraints described in Eqs.~(\ref{C1}, \ref{C2}, \ref{C3}).
The scaling functions (\ref{fpm}) are shown as lines in Figs.~\ref{fig1} 
and~\ref{collapse}.

We have noted several times the similarity between the 
present theoretical results and the numerical results 
and analysis in Refs.~\cite{tom,tom2}. In the simulations,
a scaling analogous to the one in Fig.~\ref{collapse}
was presented for the behaviour of $\log \tau_\alpha(\phi,T)$ 
instead of $\tau_\alpha(\phi,T)$ here. This implies that 
the scaling behaviour predicted by MCT is in fact in strong 
quantitative disagreement with numerical results. 
This should not come as a surprise since MCT is not able to 
describe the thermally activated relaxation which takes 
place in real glass-formers. Thus, MCT 
predicts algebraic divergences which are never observed 
in simulations and experiments, and are replaced 
by stronger, generically exponential, divergences.

It is interesting to note that algebraic scaling behaviours
and divergences seem to be well obeyed in the case of 
non-equilibrium driven athermal dynamics studied in 
Refs.~\cite{durian,teitel,hatano}. This is again 
not surprising since in these dynamics the system simply relaxes
to the nearest energy minimum without being able to cross 
energy barrier using thermal activation~\cite{jorge}. This suggests
that mode-coupling approaches and `mean-field' models (see, \textit{e.g.}
Ref.~\cite{mari}) might well be 
excellent starting points to tackle the athermal driven dynamics
of soft repulsive spheres. 
 
\section{Phase diagram}
\label{phasediag}

In this section, we move from scaling properties very near
$\phi_c$ and give a broader perspective on the behaviour 
of the system in the $(\phi,T)$ phase diagram. 
The scaling results presented above suggest 
that the system is ergodic at all temperatures 
when $\phi \le \phi_c$. For $\phi \gtrsim \phi_c$, 
Eq.~(\ref{fpm}) predicts
that the relaxation time diverges
at $\tau_\alpha \sim (T-T_c(\phi))^{-\gamma^{\rm HS}}$
with a critical temperature which vanishes
continuously at $\phi_c$ as:
\be
T_c(\phi \ge \phi_c) \sim (\phi - \phi_c)^2.
\label{tc}
\ee
Since the scaling behaviour in Eq.~(\ref{scaling}) 
only holds up to $\phi \approx 0.53$ we have fitted the temperature 
dependence of the relaxation time at larger volume fraction 
to the usual power law divergence found within MCT, 
\be
\tau_\alpha(\phi,T) \sim (T - T_c(\phi))^{-\gamma(\phi)},
\label{mctfit}
\ee
using $T_c(\phi)$ and $\gamma(\phi)$ as fitting parameters.

\begin{figure}
\includegraphics[width=8.5cm]{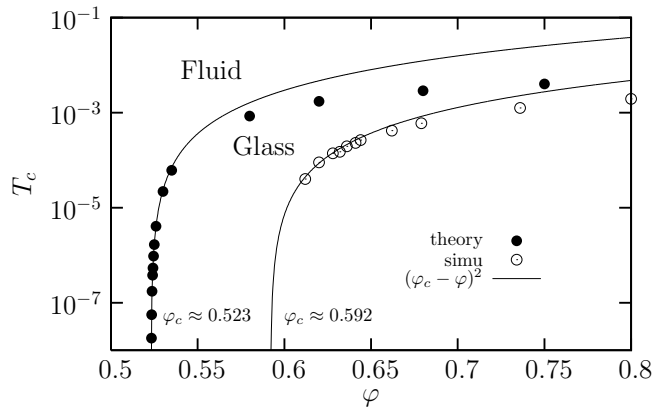}
\caption{\label{fig2} Phase diagram obtained from the MCT analysis.
Filled circles are transition temperatures 
obtained in this work, which follow the scaling behaviour 
in Eq.~(\ref{tc}) in the vicinity of $\phi_c$, as shown 
by the full line. Open symbols are the mode-coupling 
critical temperatures obtained in Ref.~\cite{tom}.}
\end{figure}

In Fig.~\ref{fig2} we show the evolution of the resulting
$T_c(\phi)$ which thus delimits the fluid and glass phases
in the theoretical phase diagram of the system. These data confirm
that the scaling behaviour in Eq.~(\ref{tc}) of the critical temperature 
is only obeyed in the vicinity of $\phi_c$, and clear deviations 
are seen at larger volume fractions where the scaling prediction
overestimates $T_c$ by quite a large amount.   

We also find that increasing the volume fraction affects 
the value of the critical exponent $\gamma$. While 
$\gamma(\phi \approx \phi_c)  = \gamma^{\rm HS} \approx 3.26$, 
we find that $\gamma$ decreases rapidly 
when $\phi$ increases: $\gamma(0.58) \approx 2.92$, 
$\gamma(0.62) \approx 2.71$, 
$\gamma(0.68) \approx 2.50$, 
$\gamma(0.75) \approx 2.41$, 
and  $\gamma(0.90) \approx 2.37$
This is a clear indication that 
when moving away from $\phi_c$ the system 
also leaves the universality class of the hard sphere 
transition. This means that 
it becomes impossible to describe the soft spheres 
as `renormalized' hard spheres when $\phi$ becomes 
too large.

\begin{figure}
\psfig{file=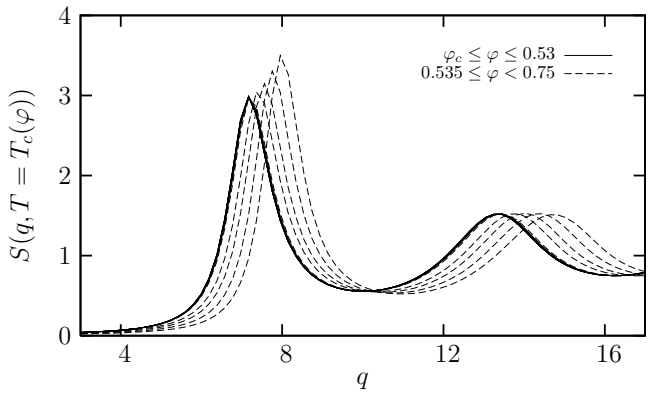,width=8.5cm}
\caption{\label{sq} Evolution of the static structure 
factor along the MCT critical line $T_c(\phi)$.
Full lines show that volume fractions between $\phi_c = 0.52315$
and $\phi=0.53$ collapse on the hard sphere structure factor 
at $\phi_c$, while deviations appear and increase rapidly
at larger $\phi$ showing that soft spheres are not simply 
`renormalized' hard spheres far above $\phi_c$.
Volume fractions are: (i) 0.5234, 0.5235, 0.5237, 0.524, 
0.5245, 0.525, 0.526, 0.5275, 0.53. (ii) 
0.535, 0.58, 0.62, 0.68, 0.75.
}
\end{figure}

We confirm  this statement in Fig.~\ref{sq} where 
we show the evolution of the static structure factor along the critical 
line $T_c(\phi)$. While nearly perfect collapse of the data
is obtained for $\phi_c \leq \phi \leq 0.53$, small deviations become 
noticeable for $\phi \approx 0.535$, and are considerably amplified 
when $\phi$ increases further. It is this large difference in the shape 
of the structure factor at the critical temperature
which accounts, within MCT, for the continuous 
evolution of the  critical
exponent $\gamma$. 

We can again compare these theoretical predictions
to the numerical analysis reported in Ref.~\cite{tom}. 
Although the mode-coupling algebraic singularity 
is not observed in real liquids, it is usually found that
such a power law behaviour is obeyed over a limited 
time window of approximately 3 decades, which allows a rough determination
of the location of the `avoided' mode-coupling 
singularity. The outcome of this exercise for the dynamics of harmonic spheres 
as determined in computer simulations reported in Ref.~\cite{tom} is 
shown in Fig.~\ref{fig2} with open symbols. The 
mode-coupling line determined numerically 
has qualitatively the same behaviour as the theoretical
line. Although the data do not span a very large 
temperature window, they are indeed compatible 
with a power law scaling as in Eq.~(\ref{tc})
near the hard sphere mode-coupling singularity 
$\phi_c$, and the critical line becomes smaller than the 
scaling prediction at larger density. 
However, the theory is quantitatively 
inaccurate as it significantly overestimates
the critical temperatures at all $\phi$. Note that for hard spheres 
MCT combined with HNC structure factor
underestimates $\phi_c$ by an amount comparable to that
reported for MCT combined with PY structure factor. These discrepancies
are well-known features of the mode-coupling approach~\cite{gotze},
and MCT descriptions of real data usually imply analysis
of scaling behaviour near singularities whose 
locations must be self-consistently determined by fitting.

A more surprising disagreement between theory and simulations
is the evolution of the critical exponent with density. 
While theory predicts a substantial decrease
of $\gamma$ at large volume fraction, numerical 
results indicate that $\gamma$ increases instead very 
rapidly with $\phi$ above the hard sphere value~\cite{tom}. 
It is not clear whether this disagreement stems from 
an incorrect prediction of the liquid structure 
by the HNC closure, or from the mode-coupling approach itself.

\section{Dynamic heterogeneity}
\label{chi4}

A newer and lesser known application of MCT is to use it to estimate  the
strength of dynamic heterogeneity accompanying the glass transition
using multi-point dynamic susceptibilities~\cite{BB}. 
It is well-known that dynamics near the glass transition 
is spatially heterogeneous, meaning that different parts of 
the system relax at different rates, while relaxation 
is correlated over a lengthscale which increases
when the glass transition is approached~\cite{ediger}. 
 
A useful tool to quantify the strength of dynamic
heterogeneity is the four-point dynamic susceptibility 
$\chi_4(t)$ which is defined from the spontaneous 
fluctuations of time correlation functions~\cite{franz,toni}:
\be
\chi_4(t) = N [ \langle f^2(q;t) \rangle - \langle f(q;t) \rangle^2 ],
\ee 
where $f(q;t)$ represents the instantaneous value of 
the intermediate scattering function $F(q;t)$. 
Intuitively, $\chi_4(t)$ increases if correlations
within the system get large, as the number of independently
relaxing units within the sample decreases~\cite{mayer}.
Formally, $\chi_4(t)$ is also the volume integral
of a spatial correlator quantifying the 
extent of correlations between local, spontaneous 
fluctuations of the dynamics and can thus directly
be considered as a proxy for the number of particles
that relax in a correlated manner close to the glass 
transition~\cite{dalle}.
  
We build on the results of 
Refs.~\cite{science05,jcpI,jcpII}
and estimate the spontaneous dynamical fluctuations 
quantified by $\chi_4(t)$ using linear response theory: 
\be
\chi_4 (t) \simeq \frac{T^2}{c_V}  \left( \frac{\partial F(q;t)}{\partial T}
\right)^2 + S(0,T) \phi^2 \left( \frac{\partial F(q;t)}{\partial \phi}
\right)^2.
\label{chi4proxy}
\ee
This relation is known to be an accurate 
representation of $\chi_4(t)$ within the 
MCT approach~\cite{jcpI} and amounts to measuring the 
response of the averaged dynamics to external fields
in the linear regime~\cite{science05}.  

The expression in Eq.~(\ref{chi4proxy}) 
is highly convenient in the present context
as we can directly obtain analytical results for the scaling 
behaviour of $\chi_4(t)$ in the vicinity of $\phi_c$ using 
results from the previous sections. Since we are  interested
in the scaling properties of the dynamic susceptibility, we 
make two further approximations to obtain an analytical 
form. We first use the fact that the time decay of the intermediate
scattering function obeys time temperature superposition, 
$F(q;t) \simeq {\cal F}(t / \tau_\alpha)$.  
Thus we have (with $x = T, \phi$):
\be
\frac{\partial F(q;t)}{\partial x} = -\frac{t}{\tau_\alpha} 
{\cal F}' \left( \frac{t}{\tau_\alpha} \right) 
\frac{\partial \ln (\tau_\alpha)}{\partial x} 
\equiv \chi_x(t),
\ee 
which is a non-monotonic function of time with a maximum 
for $t \approx \tau_\alpha$. We focus on the height of this maximum,
$\chi_4 \equiv \chi_4(t=\tau_\alpha)$,
which can then be estimated from the 
behaviour of the relaxation time alone: 
\be 
\chi_4 = C^2 \left[ 
\frac{T^2}{c_V} \left( \frac{\partial \ln \tau_\alpha}{\partial 
T} \right)^2 + S(0,T) \phi^2 \left( \frac{\partial \ln 
\tau_\alpha}{\partial \phi} \right)^2 \right], 
\label{chi4analytic}
\ee
with $C = {\cal F}'(1)$, so that $C = \beta e^{-1}$ for 
a stretched exponential lineshape,
${\cal F}(x) \sim \exp( - x^\beta)$.

To proceed analytically we make 
use of the scaling form in Eq.~(\ref{scaling}) for the 
relaxation time. We get:
\be
\chi_4 / C^2 =  \frac{1}{4 c_V} G_{\pm}^2(x)
+ S(0,T) \left( \frac{\gamma^{\rm HS} \phi}{| \delta \phi|} \right)^2 
( 1 - G_{\pm}(x)/\gamma^{\rm HS})^2,
\label{chi4scal}
\ee 
where $G_{\pm}(x) \equiv x f_{\pm}'(x)/f_{\pm}(x)$.

\begin{figure}
\includegraphics[width=8.5cm]{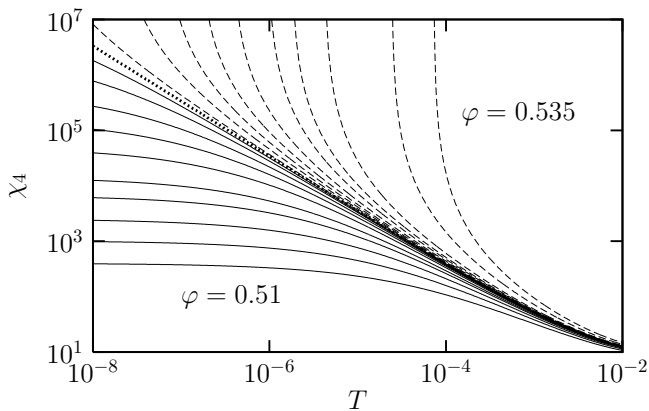}
\caption{\label{fig3} Evolution of the peak of 
the dynamic susceptibility $\chi_4$ estimated from 
Eq.~(\ref{chi4scal}) for the same volume fractions 
of in Fig.~\ref{fig1}, including data below (full
lines), at (dotted line), and above (dashed lines)
$\phi_c$. The behaviour is clearly reminiscent of the one 
of $\tau_\alpha(\phi,T)$ in Fig.~\ref{fig1}.}
\end{figure}

In Fig.~\ref{fig3} we show the evolution of $\chi_4$,
evaluated from Eq.~(\ref{chi4scal})  across $\phi_c$. 
It is clear from this figure that $\chi_4$ has 
scaling properties very similar to the ones of
the relaxation time (Fig.~\ref{fig1}). 
It increases and saturates to a plateau when 
$T$ decreases for $\phi < \phi_c$, obeys a power law behaviour 
at $\phi_c$, and diverges algebraically 
at $T_c(\phi)$ above $\phi_c$. 

In particular, we find 
that the term proportional to $c_V^{-1}$ 
and stemming from the temperature derivative in Eq.~(\ref{chi4proxy})
is always safely negligible to the volume fraction derivative term 
in the dynamic range shown in Fig.~\ref{fig3}.
In fact, this figure would be almost unchanged if we 
had shown only the second term 
in Eq.~(\ref{chi4proxy}).
Physically this implies, not too surprisingly, 
that dynamic heterogeneity in the scaling regime of the harmonic
sphere system is mainly controlled by density 
fluctuations, just as for hard spheres~\cite{luca}, while 
energy fluctuations play little role. We note that 
the opposite is true in supercooled liquids, 
where density fluctuations seem to be generically dominated 
by energy fluctuations~\cite{dalle,jcpI}. 

A second interesting consequence is that the scaling 
behaviour of $\chi_4$ near $\phi_c$ can then 
be obtained analytically: 
\be
\chi_4 \approx \chi_4^{\rm HS}(\phi) {\cal X}_{\pm} \left( \frac{|\delta 
\phi|}{\sqrt{T}} \right),
\label{scalchi4}
\ee 
where 
\be
\chi_4^{\rm HS}(\phi) \sim \phi^2/(\phi_c - \phi)^{2},
\label{chi4hs}
\ee 
is the
hard sphere result~\cite{jcpI}, and 
${\cal X}_{\pm}(x) = (1-G_{\pm}(x)/\gamma^{\rm HS})^2$.
The scaling behaviour in Eq.~(\ref{scalchi4}) is 
similar to the one 
of $\tau_\alpha(\phi,T)$ found in Eq.~(\ref{scaling}). 
In fact the similarity is even quantitative, since combining
Eqs.~(\ref{scaling}, \ref{fpm}, \ref{scalchi4}), we can 
explicitly show that the relationship between 
the four-point susceptibility and the averaged
relaxation time is identical for soft and hard spheres 
in the scaling regime near $\phi_c$, up to subleading contributions.
This suggests that plotting $\chi_4(\phi,T)$ vs. 
$\tau_\alpha(\phi,T)$ would collapse
all data for harmonic spheres onto the hard sphere 
data.

\begin{figure}
\includegraphics[width=8.5cm]{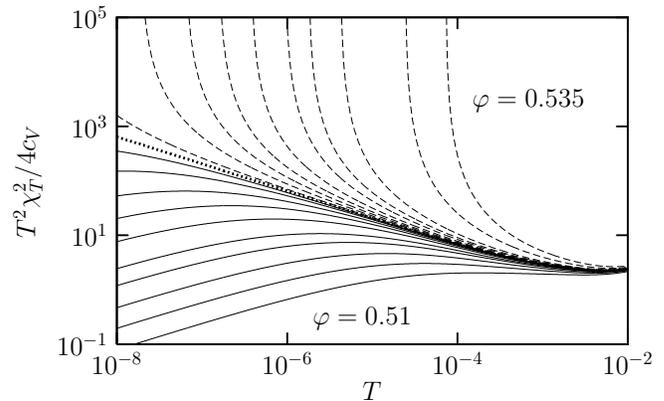}
\caption{\label{fig3b}  
Evolution of the contribution of the term containing the 
temperature in  Eq.~(\ref{chi4proxy}) for the same parameters and using 
the same representation as in Fig.~\ref{fig3}.
Note that the vertical scale in both figures is different, and 
that the thermal contribution to $\chi_4$ is always
much smaller than the density contribution.}
\end{figure}

For completeness we also show the `thermal' contribution to
$\chi_4(t)$ in 
Eq.~(\ref{chi4proxy}) in
Fig.~\ref{fig3b}, because this term is too small
to have observable effects in Fig.~\ref{fig3}. 
Although its shape seems similar to the one of 
$\chi_4$, it is not quite the same: it vanishes as $T \to 0$ 
for $\phi < \phi_c$, because $\tau_\alpha$ does not depend 
on $T$ in this limit. It follows a power law behaviour 
for $\phi = \phi_c$, but this divergence is in fact 
entirely due to the $1/c_V$ prefactor, since the specific heat 
behaves as $c_V(T) \sim \sqrt{T}$ from Eq.~(\ref{T32}). Finally, 
for $\phi > \phi_c$ both terms contributing 
to $\chi_4$ diverge in the same power law manner, 
as $(T-T_c(\phi))^{-2}$, but the respective amplitude 
of the two terms is set by $(\phi/\delta \phi)^2$ and $1/c_V$.
This implies that as long as $\phi$ is close to $\phi_c$ the density
derivative term dominates over the temperature derivative 
term. It is only when 
$\phi$ is much larger than $\phi_c$ that the temperature contribution 
might become dominant, but $\chi_4$ is not described by 
Eq.~(\ref{chi4scal}) anymore and a direct evaluation of 
all contributions would be required to investigate this
crossover at large volume fractions. 
We have not pursued these investigations.  

To the best of our knowledge, dynamic heterogeneity has not been 
discussed numerically in harmonic spheres, and we cannot 
compare the present results with numerical results. 
However, for hard spheres, it has been established 
that the power law scaling in Eq.~(\ref{chi4hs}) is barely visible 
on actual data~\cite{luca}, and dynamic correlations seem to increase 
much more slowly with increasing the density than 
predicted by MCT, as is found also for 
supercooled liquids~\cite{dalle}. 

We note again, however, the close similarity between the
present MCT results for dynamic correlations and the 
scaling properties found in driven athermal simulations of harmonic
spheres where algebraic divergences of spatial correlations 
of particle dynamics and scaling properties very similar to
Fig.~\ref{fig3} were reported~\cite{teitel,hatano,hatano2}.
 
\section{Discussion}
\label{conclusion}

In this paper, we have investigated theoretically 
the behaviour of dense assemblies of harmonic spheres 
at low temperatures in a broad range of volume fractions, 
encompassing the glass transition of hard spheres at $\phi_c$. 
We have combined 
hyper-netted chain closure for the structure with mode-coupling 
theory for the dynamics. We find that for finite temperatures
near $\phi_c$, harmonic spheres behave effectively as 
hard spheres with a renormalized volume fraction. This directly
implies a scaling form for the relaxation time in the part of the 
volume fraction -- temperature phase diagram in the vicinity of the
hard sphere transition, which applies
also to the amplitude of dynamic heterogeneity.  At larger volume 
fraction, deviations from hard sphere behaviour arise, and 
dynamic scaling breaks down.

When compared to numerical simulations of the dynamics
of harmonic spheres at thermal equilibrium, the known shortcomings
of mode-coupling predictions clearly show up: MCT predicts 
algebraic singularities that are not observed in simulations, 
and fails to predict the `activated' scaling behaviour 
observed in simulations of both hard and soft particles.     
In particular, using a mode-coupling approach we cannot
discuss the large change of glass fragility with volume 
fraction discussed in Refs.~\cite{tom,tom2}, as 
these come from subtle deviations from an Arrhenius behaviour
which is not predicted with MCT. Another feature 
which is not well captured by the present calculations 
is the volume fraction dependence of the MCT critical exponent which 
increases with $\phi$ in the simulations, but decreases   
in our calculations. It is not clear, however, whether this last failure
originates from the approximation used for the structure factor or
from the mode-coupling theory itself.

Although MCT is qualitatively unable to describe the 
nature of the glass transition in harmonic spheres, 
there seems to exist a time window of approximately 3 decades 
where its predictions can be applied. In experiments
with hard sphere colloids, it is in fact only very recently
that deviations from MCT behaviour were  
unambiguously observed in experiments covering a very 
broad range of relaxation times~\cite{luca}. This means 
in turn that for `standard' experiments
the behaviour predicted in the present article might 
still be of some value, and our theoretical approach 
could certainly be extended to a broader family of 
soft pair potentials beyond harmonic interactions.
Thus, we hope that the present results will motivate further
analysis of the dynamics of soft colloids in 
simulations and experiments. 

Although we mentioned in the introduction that soft colloids
are currently studied by several groups, the glass transition 
of soft colloids has only very recently been studied in a system
made of microgel particles~\cite{weitz}. 
In this paper, three types of particles 
with increasing softness were studied. The most striking result
of this study is a change of the volume fraction 
dependence of the relaxation time with softness, from 
$\tau_\alpha \sim (\phi_c - \phi)^{-\gamma^{\rm HS}}$ 
for hard particles, to $\log (\tau_\alpha) \propto \phi$ for very soft 
particles. Since the interparticle interaction in this system
is not known, one could imagine using a potential such as 
in Eq.~(\ref{pair}), with increasing temperature playing the role of 
increasing particle softness. Our results in fact predict that 
the volume fraction dependence of the relaxation time does not vary 
from the hard sphere behaviour for 
a very broad range of temperatures of at least 6 decades, 
see Fig.~\ref{fig1}. Since the qualitative behaviour 
found in this work should be  independent of the 
details of the pair potential, 
this suggests that, very likely, the change of 
particle softness in Ref.~\cite{weitz} also corresponds
to a change of the form of the interaction between particles,
and is thus difficult to explain theoretically
on the basis of the present work.

Does the scaling behaviour found for harmonic spheres 
teach some lessons for understanding the glass transition 
of molecular glass-formers? A major conclusion drawn from 
the present theoretical results is that the physics
of harmonic spheres is simply the one of 
hard spheres: they undergo a glass transition 
both upon compression or upon cooling with a relaxation time which 
diverges when the effective volume fraction $\phi_{\rm eff}(\phi,T)$ of
the harmonic spheres system becomes equal to $\phi_c$,
the hard sphere critical packing fraction. Moreover, the physics 
for $\phi_{\rm eff}(\phi,T) \lesssim \phi_c$ is the same for both soft
and hard particles, as seen for instance from the behaviour
of the dynamic susceptibilities discussed in Sec.~\ref{chi4}.  

For this physical behaviour to be useful to understand
aspects of the glass transition, one should invoke 
the possibility that real liquids can be effectively 
described as hard spheres, an assumption which has a long history in 
the field of liquid state theory~\cite{barrat}. 
It was revisited very recently in 
the present context in Ref.~\cite{gilles} which 
established that the dynamics of real liquids, 
and in particular the interplay between density and temperature,
is qualitatively different from the one of soft particles
with a finite interaction range such as harmonic spheres. 
For instance the large change of glass fragility observed for 
harmonic spheres is not observed in molecular liquids,
which obey a much simpler scaling yielding glass fragilities
independent of the density~\cite{gilles}. 
Here, we also found that the main contribution 
to the dynamic susceptibility in Eq.~(\ref{chi4proxy}) is always
given by the density contribution, while in real 
liquids the temperature term dominates~\cite{jcpI,jcpII,dalle}, 
suggesting that 
a different physics is at play in both cases. It could be, 
for instance, that attractive 
forces not included in potentials such as Eq.~(\ref{pair}) provide a 
non-negligible contribution to the energy barriers 
that need to be crossed during 
structural relaxation.

Finally, we comment on the intriguing similarity emphasized 
throughout this paper between 
the present results and the dynamic scaling
behaviour discussed in several recent articles dealing 
with the athermal, driven dynamics of harmonic 
spheres~\cite{teitel,hatano,hatano2}. In both cases,
algebraic divergences of dynamical quantities are obtained 
in the hard sphere limit, with a dynamic scaling behaviour 
observed in its vicinity both at the level of the averaged 
dynamics and of the dynamical fluctuations.
Note, however, the very different nature of the critical 
density in both cases~\cite{jorge}: the hard sphere glass transition 
discussed here is defined from the divergence of an
equilibrium quantity, which is thus by definition 
independent of the preparation protocol of the system.
Instead the zero temperature jamming transition 
does explicitly depend on which ensemble of configurations
is selected by the studied dynamics~\cite{ohern,mari,torquato,pinaki}.
There is no limit where these two distinct transitions can merge.
 
We believe  that the deep underlying explanation of this similarity
is the fact that in both cases, the dynamics is controlled 
by the existence of `soft modes', meaning that relaxation 
proceeds both within MCT and in athermal dynamics without 
spontaneous crossing of energy barriers. In both cases, thus,
nontrivial collective dynamics stems from the existence 
of nearly flat directions of the potential 
energy landscape~\cite{jorge,hatano2}. 
This is a clear shortcoming of MCT when it deals 
with the glass transition of glass-forming liquids, 
but since no such barrier crossing takes place in the $T=0$
driven dynamics relevant for granular systems,
we suggest that `mean-field' mode-coupling approaches
such as the ones developed here and by others~\cite{mari} 
provide good starting points
to describe the dynamics of soft particles near the jamming 
transition at zero temperature.  
 
\acknowledgments
We thank L. Cipelletti, W. Kob, G. Tarjus, T. Witten, 
F. Zamponi for useful exchanges about this work. 
This work was started when G. Szamel was on sabbatical leave at LCVN.
He is grateful to his colleagues there for their hospitality 
and to CNRS for financial support
that made his stay in Montpellier possible. G. Szamel and E. Flenner 
gratefully acknowledge the support of NSF Grant No.\ CHE 0517709.
H. Jacquin acknowledges financial support from Capital Fund Management (CFM)
Foundation, and from the LCVN in the early stages of this work.
L. Berthier is partially funded by ANR Dynhet.

\end{document}